\documentclass[10pt]{sensys-abstract}

\usepackage{graphicx}
\graphicspath{{fig/}}
\usepackage{cite}
\usepackage{xspace}

\newcommand{\etal}{et~al.\xspace}
\newcommand{\eg}{e.g.,\xspace}
\newcommand{\ie}{i.e.,\xspace}

\newcommand\figref[1]{Figure~\ref{#1}}
 
\newcommand\secref[1]{Section~\ref{#1}}

\permission{Copyright is held by the author/owner(s).}
\conferenceinfo{SenSys'10,} {November 3--5, 2010, Zurich, Switzerland.}
\CopyrightYear{2010}
\crdata{978-1-4503-0344-6/10/11}

\begin{document}

\title{Poster Abstract: If You Have Time, Save Energy with Pull}

\numberofauthors{1}
\author{
\alignauthor David Hasenfratz, Andreas Meier, Matthias Woehrle, Marco Zimmerling, Lothar Thiele\\
\affaddr{Computer Engineering and Networks Lab, ETH Zurich, Switzerland}\\\email{hdavid@ee.ethz.ch \hspace{5pt} \{a.meier,woehrle,zimmerling,thiele\}@tik.ee.ethz.ch}}

\maketitle

\begin{abstract}
We analyze push and pull for data collection in wireless sensor networks. Most applications to date use the traditional push approach, where nodes transmit sensed data immediately to the sink. Using a pull approach, nodes store the data in their local flash memory, and only engage in communication during dedicated collection phases. We show how one can transform an existing push-based collection protocol into a pull-based one, and compare the power consumption of both approaches on a 35-node testbed. Our results show that substantial energy gains are possible with pull, provided that the application can tolerate a long latency.
\end{abstract}

\section{Introduction}

Data collection is the mainstream application scenario in wireless sensor networks~(WSNs), where sensor nodes collect and forward data to a base station for further processing and analysis. Since the nodes are usually battery-powered, energy efficiency is a major requirement for long-term data collection. While latency plays only a minor role in data collection, a data yield close to 100$\,$\% is typically expected.

\begin{figure}
 \centering
 \includegraphics[width=\columnwidth]{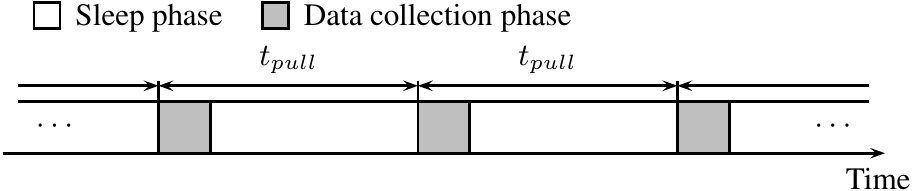}
 \caption{Basic structure of pull-based data collection.}
 \label{fig:pull}
\end{figure}

In this paper, we analyze two different approaches for energy-efficient data collection in WSNs: push and pull. Using the \emph{push} approach, nodes immediately transmit the sensed data to the sink. This requires the nodes to be continuously active to maintain up-to-date routing paths. The \emph{pull} approach instead operates in two alternating phases, as illustrated in \figref{fig:pull}. During regular sleep phases the nodes do not transmit sensed data but store these in the local flash memory. At the beginning of each data collection phase, the sink wakes up the network, whereupon the nodes transmit all data backlogged in the preceding sleep phase. Afterwards, the sink puts the network back to sleep.

The major advantage of push is the lower latency. Its downside is the energy overhead to maintain up-to-date routing paths. This is not required in the pull approach, which allows the nodes to limit their activity to a minimum in the sleep phase to save energy. These savings are paid by i)~control overhead incurred by the central administration, ii)~energy overhead due to flash data storage and routing path initialization before each data collection phase, and iii)~higher latency as determined by the pull interval $t_{pull}$.

Most data collection applications use push, even though the application requirements would allow for pull (\eg no latency requirement). Dutta~\etal~\cite{Dutta2007} present first insights into a pull-based approach. We build on these promising results by comparing the energy consumptions of push and pull based on testbed experiments. We also show how one can modify an existing, push-based data collection protocol to make it work like a pull-based one. These modifications are applicable to any push-based protocol running on top of any medium access control~(MAC) protocol based on low power listening. We demonstrate the feasibility of our techniques with a basic implementation of the Collection Tree Protocol~(CTP)~\cite{Gnawali2009b} running on top of the X-MAC~\cite{Buettner2006} protocol.

\section{Making it Work}\label{sec:approach}

To transform a push-based data collection protocol such as CTP into a pull-based one, we start by splitting its operation in two distinct phases: a sleep phase and a data collection phase. These phases must satisfy the following conditions:
\begin{itemize}
 \item Sleep phase: Nodes do not communicate while waiting for the sink to initiate the data collection phase. Sensed data is stored locally in flash memory.
 \item Data collection phase: Nodes first initialize the routing paths as done when starting the push-based data collection. As soon as routing paths are available, nodes start to transmit backlogged data.
\end{itemize}

In a second step, we leverage the beacons periodically broadcast by the routing protocol to exchange routing information for the central administration of pull. Specifically, we extend these beacons with phase IDs to distinguish the two protocol phases. To trigger a phase change, the sink sets the appropriate phase ID in its beacons.

The energy consumption in the sleep phase is heavily dictated by the idle listening of the MAC protocol. While a long MAC wake-up interval $T_w$ saves energy in the sleep phase, it limits the available bandwidth in the data collection phase. To overcome this limitation, we propose to use two different wake-up intervals: $T_w^s$ in the sleep phase and $T_w^c$ in the data collection phase. The longer $T_w^s$ saves energy by achieving a low duty cycle in the sleep phase. The shorter $T_w^c$ provides sufficient bandwidth in the data collection phase.

For a fair comparison between push and pull with respect to energy, we perform two additional modifications to achieve close to 100$\,$\% data yield for both approaches:
\begin{itemize}
 \item A data packet is only deleted from the sender's buffer after the reception of a data packet acknowledgment.
 \item If the receiver's forwarding buffer~\cite{Gnawali2009b} is full, it replies with an explicit ``not acknowledged'' message. This forces the sender to back off before sending the data packet again, and hence works as a back-pressure flow-control mechanism.
\end{itemize}

\section{Experimental Comparison}

\begin{figure}
 \centering
 \includegraphics[width=\columnwidth,viewport=50 0 625 300,clip]{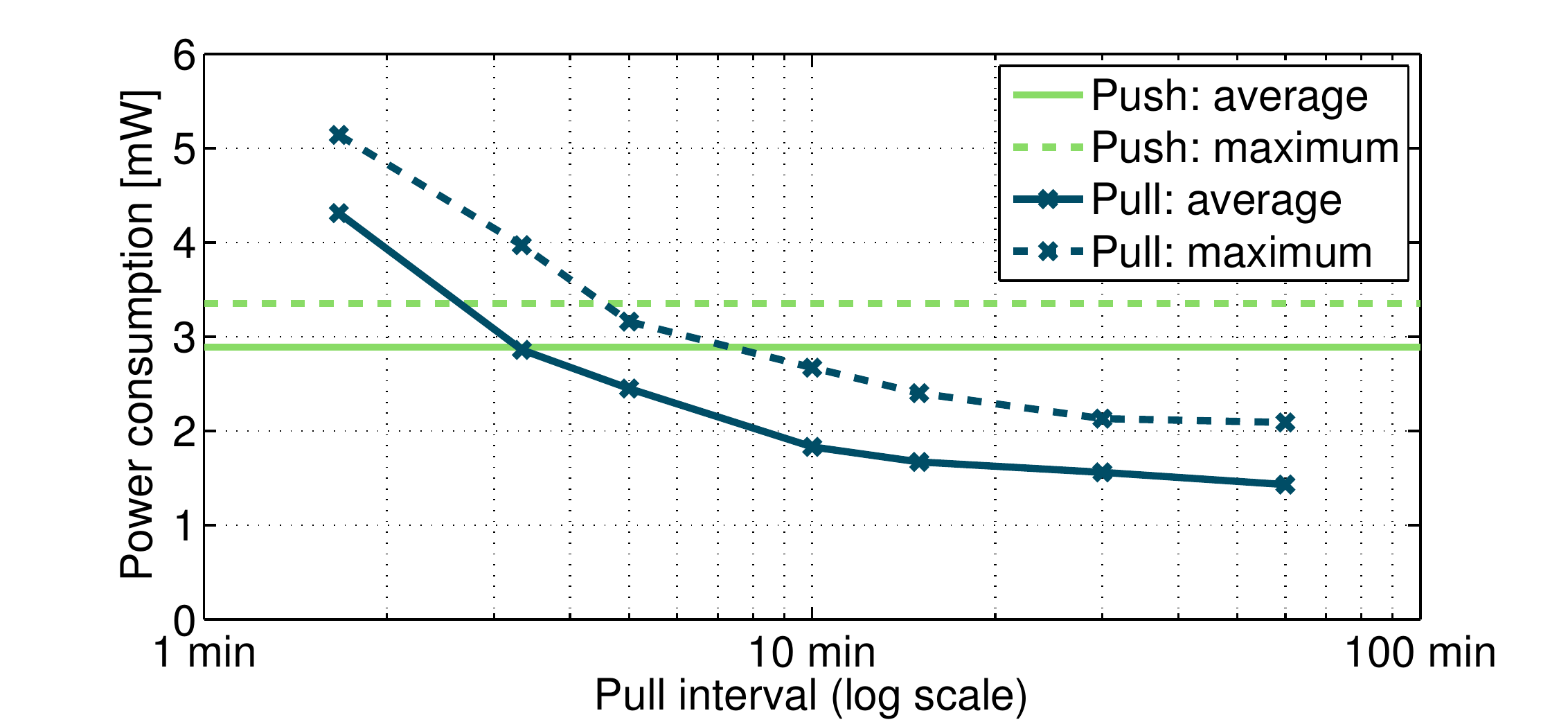}
 \caption{Power consumption comparison of the push and pull approach for varying pull intervals.}
 \label{fig:pull_int}
\end{figure}

We compare the power consumptions (\ie microcontroller, radio transceiver, and flash) of push and pull on a 35-node testbed. The nodes are Tmote Sky~\cite{Polastre05} devices distributed in several offices on one floor, running the Contiki operating system~\cite{Dunkels2004}. Nodes generate data with a sampling interval of 45~seconds. One node acts as an always-listening sink, \ie its radio is always on. We do not consider the power consumption of the sink as it has an unlimited energy supply. We receive with both approaches in all experiments 100$\,$\% of the generated packets, due to the reliability modifications described in \secref{sec:approach}.

We first analyze the power consumption of pull for intervals $t_{pull}$ ranging from 100~seconds to 60~minutes and compare these with push. This is shown in Figure~\ref{fig:pull_int}. We look at the average and maximum power consumption. Average denotes the average power consumption of all nodes and maximum is the maximum power consumption among all nodes.

Every data collection phase incurs an energy overhead for waking up the network and putting it back to sleep. The fewer data collection phases are performed, i.e., the longer the pull interval, the lower the total energy consumption. 
Pull uses less energy than push with pull intervals above 5~minutes. For example, with a pull interval of 60~minutes, the average and the maximum power consumptions of pull are 50$\,$\% and 40$\,$\% below those of push.

\begin{figure}
 \centering
 \includegraphics[width=\columnwidth,viewport=23 0 555 280,clip]{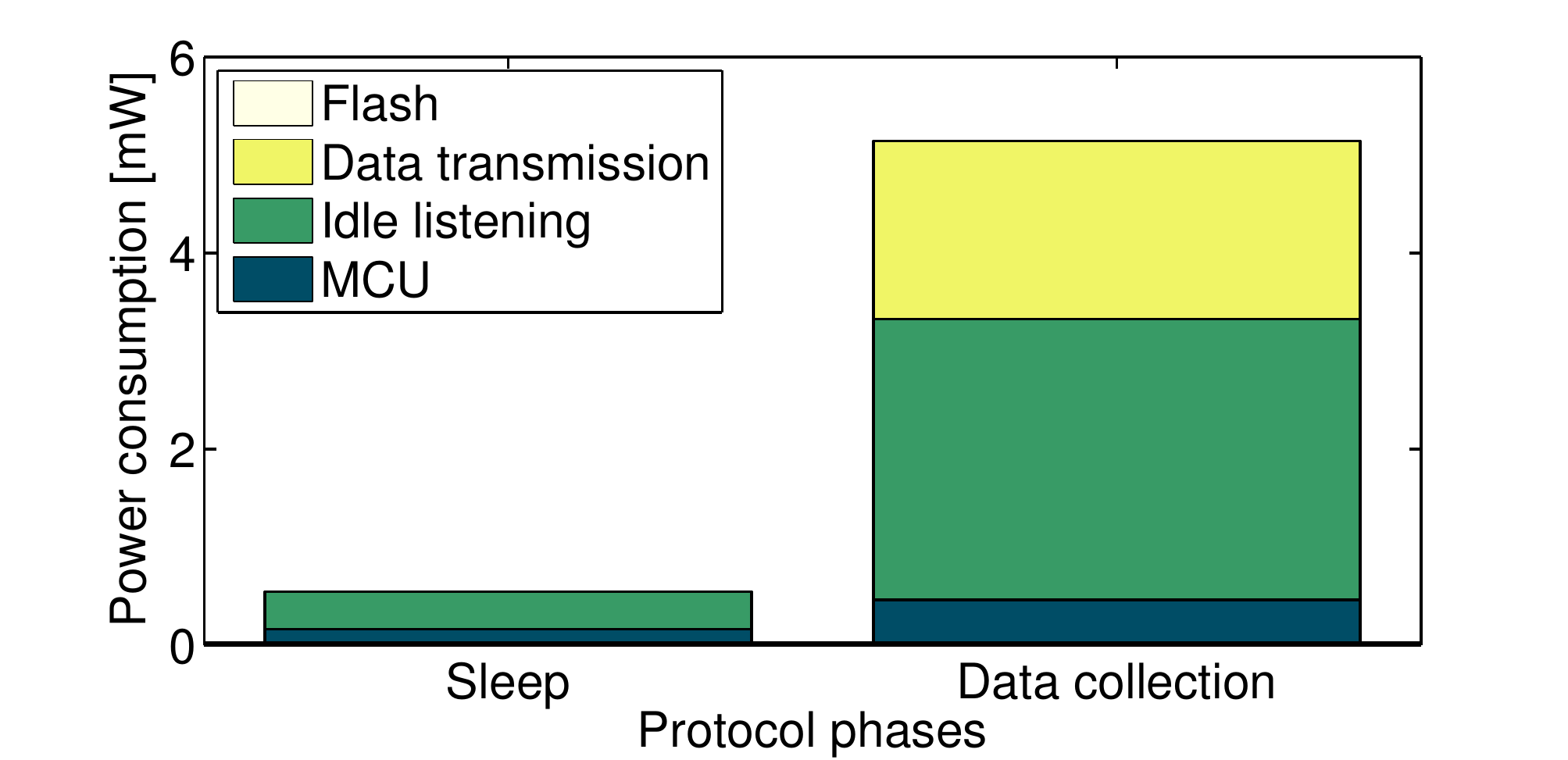}
 \caption{Average power consumption of pull in the sleep and the data collection phases.}
 \label{fig:pull_power_bar}
\end{figure}

The distribution of power consumption for the pull approach with a pull interval of 60~minutes is depicted in Figure~\ref{fig:pull_power_bar}. It shows the average power consumptions of the microcontroller~(MCU), the radio (for idle listening and data transmission), and the flash in the two phases. Nodes consume 90$\,$\% less power in the sleep phase compared to the data collection phase, because there is no data transmission and especially due to the longer MAC wake-up interval. Notably, the power used by the flash is negligible in both phases.

\section{Conclusions}

Even using a simple pull approach achieves an energy gain of almost a factor of two. The energy overhead due to recurring initialization phases is amortized if the pull interval is above a couple of minutes. This is because the energy saved in the sleep phases is much larger. With respect to energy consumption it is favorable to have as few collection phases as possible. The maximum pull interval is bounded by the latency requirement and the available flash storage. The increasing amount of available flash storage relaxes the storage requirement. For example, having just 1~MB flash storage and sampling 100 Bytes of data once per minute would require to collect data only once a week. Furthermore, the pull approach provides a wide range of power optimization opportunities~\cite{Dutta2007}, such as efficient data stream transmissions and effective data compression due to the large amount of available data.

\section{Acknowledgements}
\label{sec:acknowledgements}
The work presented here was supported by the National Competence Center in Research on Mobile Information and Communication Systems (NCCR-MICS), a center supported by the Swiss National Science Foundation under grant number 5005-67322.

\bibliographystyle{abbrv}
{\small
\bibliography{biblio}

\begin{thebibliography}{1}

\bibitem{Buettner2006}
M.~Buettner, G.~V. Yee, E.~Anderson, and R.~Han.
\newblock {X-MAC}: A short preamble {MAC} protocol for duty-cycled wireless
  sensor networks.
\newblock In {\em Proc. of SenSys '06}.

\bibitem{Dunkels2004}
A.~Dunkels, B.~Gr\"onvall, and T.~Voigt.
\newblock Contiki -- a lightweight and flexible operating system for tiny
  networked sensors.
\newblock In {\em Proc. of LCN~'04}.

\bibitem{Dutta2007}
P.~Dutta, D.~Culler, and S.~Shenker.
\newblock {Procrastination might lead to a longer and more useful life}.
\newblock In {\em Proc. of HotNets-VI '07}.

\bibitem{Gnawali2009b}
O.~Gnawali, R.~Fonseca, K.~Jamieson, D.~Moss, and P.~Levis.
\newblock Collection tree protocol.
\newblock In {\em Proc. of SenSys '09}.

\bibitem{Polastre05}
J.~Polastre, R.~Szewczyk, and D.~Culler.
\newblock Telos: Enabling ultra-low power wireless research.
\newblock In {\em Proc. of IPSN '05}.

\end{thebibliography}
}

\balancecolumns

\end{document}